\documentclass[12pt,amsmath,amssymb,a4paper,final]{iopart}

\usepackage{verbatim}
\usepackage{bm}
\usepackage{dcolumn}

\usepackage{graphicx}
\usepackage{epstopdf}
\usepackage{subfigure}

\usepackage{amsfonts}
 \expandafter\let\csname equation*\endcsname\relax
   \expandafter\let\csname endequation*\endcsname\relax
\usepackage{amsmath}
\usepackage{iopams}
\usepackage{amssymb}
\usepackage{subfigure}
\usepackage{lipsum}

\usepackage{wasysym}

\usepackage{booktabs}

\usepackage{color}

\usepackage[colorlinks,bookmarks=false,citecolor=blue,linkcolor=red,urlcolor=blue]{hyperref}

\newcommand{\be}{\begin{equation}}
\newcommand{\ee}{\end{equation}}
\newcommand{\bea}{\begin{eqnarray}}
\newcommand{\eea}{\end{eqnarray}}

\newcolumntype{.}{D{.}{.}{-1}}

\def\bs{\boldsymbol}
\def\vec{\mathbf}
\def\mc{\mathcal}

\allowdisplaybreaks

\begin{document}

\title[Kitaev interactions in Na$_2$IrO$_3$]
{Kitaev interactions between $j\!=\!1/2$ moments in honeycomb Na$_2$IrO$_3$ are large and ferromagnetic:
 insights from {\it ab initio} quantum chemistry calculations}

\author{Vamshi M.~Katukuri$^1$, S.~Nishimoto$^1$, V.~Yushankhai$^{2,3}$, A.~Stoyanova$^3$,
        H.~Kandpal$^1$, Sungkyun Choi$^4$, R.~Coldea$^4$, I.~Rousochatzakis$^1$, L.~Hozoi$^1$ and Jeroen van den Brink$^{1,5}$}











\address{$^1$Institute for Theoretical Solid State Physics, IFW Dresden, Helmholtzstr.~20, 01069 Dresden, Germany}
\address{$^2$Joint Institute for Nuclear Research, Joliot-Curie 6, 141980 Dubna, Russia}
\address{$^3$Max-Planck-Institut f\"{u}r Physik komplexer Systeme, N\"{o}thnitzer Str.~38, 01187 Dresden, Germany}
\address{$^4$Clarendon Laboratory, University of Oxford, Parks Road, Oxford OX1 3PU, United Kingdom}
\address{$^5$Department of Physics, Technical University Dresden, Helmholtzstr.~10, 01069 Dresden, Germany}

\ead{v.m.katukuri@ifw-dresden.de,
s.nishimoto@ifw-dresden.de,
l.hozoi@ifw-dresden.de,
j.van.den.brink@ifw-dresden.de}

\begin{abstract}
Na$_2$IrO$_3$, a honeycomb $5d^5$ oxide, has been recently identified as a potential
realization of the Kitaev spin lattice.
The basic feature of this spin model is that for each of the three metal-metal links
emerging out of a metal site, the Kitaev interaction connects only spin components
perpendicular to the plaquette defined by the magnetic ions and two bridging
ligands.
The fact that reciprocally orthogonal spin components are coupled along the three different
links leads to strong frustration effects and nontrivial physics.
While the experiments indicate zigzag antiferromagnetic order in Na$_2$IrO$_3$,
the signs and relative strengths of the Kitaev and Heisenberg interactions are still under debate.
Herein we report results of {\it ab initio} many-body electronic-structure calculations
and establish that the nearest-neighbor exchange is strongly anisotropic with a dominant
ferromagnetic Kitaev part, whereas the Heisenberg contribution is significantly weaker and
antiferromagnetic.
The calculations further reveal a strong sensitivity to tiny structural details such as the
bond angles.
In addition to the large spin-orbit interactions, this strong dependence on distortions of
the Ir$_2$O$_2$ plaquettes singles out the honeycomb $5d^5$ oxides as a new playground for the
realization of unconventional magnetic ground states and excitations in extended systems.
\end{abstract}

\date\today


\submitto{\NJP}

\maketitle

\section{Introduction}

The Heisenberg model of magnetic interactions, $J{\bm S}_i\!\cdot\!{\bm S}_j$ between spin
moments at sites $\{i,j\}$, has been successfully used as an effective minimal model to
describe the cooperative magnetic properties of both molecular and solid-state many-electron
systems.
A less conventional spin model -- the Kitaev model \cite{Kitaev2006} -- has been
recently proposed for honeycomb-lattice materials with $90^{\circ}$ metal-oxygen-metal
bonds and strong spin-orbit interactions \cite{IrO_kitaev_jackeli_09}.
It has nontrivial topological phases with elementary excitations exhibiting Majorana
statistics, which are relevant and much studied in the context of topological quantum
computing
\cite{Kitaev2006,Kit_baskaran_07,Chen08,Vidal08,Tikhonov11,Kit_nussinov_13}.
Candidate materials proposed to host such physics are the honeycomb oxides Na$_2$IrO$_3$ and
Li$_2$IrO$_3$ \cite{IrO_kitaev_jackeli_09}.
The magnetically active sites, the Ir$^{4+}$ species, display in these compounds a $5d^5$
valence electron configuration, octahedral ligand coordination and bonding of nearest-neighbor (NN) Ir ions
through two ligands~\cite{Honeycomb_NaIrO_Choi_2012,Honeycomb_NaIrO_Ye_2012}.
In the simplest approximation, i.e., for sufficiently large $t_{2g}$--$e_g$ octahedral
crystal-field splittings within the Ir $5d$ shell and degenerate Ir $t_{2g}$ levels, the
ground-state electron configuration at each Ir site is a $t_{2g}^5$ effective $j\!=\!1/2$ spin-orbit
doublet~\cite{SOC_d5_thornley68,book_abragam_bleaney,IrO_mott_kim_08,IrO_kitaev_jackeli_09}.
The anisotropic, Kitaev type coupling then stems from the particular form the superexchange between
the Ir $j\!=\!1/2$ pseudospins takes for $90^{\circ}$ bond angles on the Ir-O$_2$-Ir plaquette
\cite{IrO_kitaev_jackeli_09,IrO_kitaev_chaloupka_10,ZigZag_KH_chalopka_12}.

Recent measurements on Na$_2$IrO$_3$ \cite{Honeycomb_NaIrO_Choi_2012, Honeycomb_NaIrO_Ye_2012}
indicate significant lattice distortions away from the idealized case of cubic IrO$_6$
octahedra and $90^{\circ}$ Ir-O-Ir bond angles for which the Kitaev-Heisenberg (KH) model was proposed
\cite{IrO_kitaev_jackeli_09,IrO_kitaev_chaloupka_10}.
Lower-symmetry crystal fields and distortions of the Ir-O-Ir bonds obviously give rise to
finite Ir $t_{2g}$ splittings \cite{213_rixs_gretarsson_2012,Mazin_Na213_2013} and more complex
superexchange physics \cite{anisotr_mgn_yushankhai99,anisotr_mgn_tornow99}.
It has been actually shown that the interplay between ``local'' distortions of the O cage
and longer-range crystal anisotropy is a key feature in $5d$ oxides
\cite{Liu_Sr3116_PRL_2012,katukuri_PRB_2012,nikolay_cairo_PRB,Ir227_hozoi_12,Os227_bogdanov_12}
and the outcome of this competition is directly related to the precise nature of the
magnetic ground state \cite{Os227_bogdanov_12}.
Moreover, the lower symmetry characterizing a given [Ir$_2$O$_{10}$] unit of two edge-sharing
octahedra allows in principle for nonzero anisotropic interaction terms beyond the
Kitaev picture.

The inelastic neutron scattering data
\cite{Honeycomb_NaIrO_Choi_2012} and the magnetic ordering pattern
\cite{Honeycomb_NaIrO_Ye_2012,Na2IrO3_zigzag_liu_2011} in
Na$_2$IrO$_3$ could in principle be explained in a minimal model
by either i) a more conventional Heisenberg model with
frustrated exchange couplings extending up to third NN's
\cite{Honeycomb_NaIrO_Choi_2012} or ii) a 
KH model with dominant antiferromagnetic (AF) Kitaev and smaller
ferromagnetic (FM) Heisenberg NN terms
\cite{ZigZag_KH_chalopka_12}. 
The presence of strong Kitaev interactions 
has furthermore been suggested on the basis of recent resonant
inelastic x-ray scattering experiments
\cite{Na2IrO3_KH_Hlynur_2013}.

To clarify the signs and the strengths of the effective coupling
constants in Na$_2$IrO$_3$, we here employ many-body {\it ab
initio} techniques from wave-function-based quantum chemistry
\cite{book_QC_00}. We establish that the situation is much more
subtle than presumed so far. In model systems the KH Hamiltonian
arises due to the destructive interference of different
superexchange pathways that contribute equally to the effective
intersite interaction
\cite{IrO_kitaev_jackeli_09,IrO_kitaev_chaloupka_10}. This
interference turns out to be rather fragile as even when we
consider idealized structures with cubic IrO$_6$ octahedra,
orthogonal Ir-O-Ir bonds and $D_{2h}$ point-group symmetry of the
Ir-Ir link, the computed low-energy magnetic spectrum does not
support a pure KH model.
A careful analysis shows that in the Kitaev reference frame \cite{IrO_kitaev_jackeli_09,IrO_kitaev_chaloupka_10}
off-diagonal terms of the symmetric anisotropic exchange-coupling tensor are allowed
by symmetry to be non-zero.
The quantum chemistry (QC) calculations predict the latter are comparable in magnitude
to the strength of the isotropic Heisenberg term.
We find that the effective Kitaev coupling is, however, the dominant energy scale
-- depending on geometrical details in the range of 10--20 meV and FM --
and that the NN Heisenberg $J$ is AF and significantly weaker.
For NN interaction parameters as derived in the QC study, we have further performed exact
diagonalization (ED) calculations including additionally finite AF second and third order
Ir-Ir Heisenberg couplings.
These indicate the presence of zigzag AF order, in agreement with the experimentally
observed spin texture \cite{Honeycomb_NaIrO_Choi_2012,Honeycomb_NaIrO_Ye_2012,Na2IrO3_zigzag_liu_2011}.

\begin{figure}[b!]
  \begin{center}
\subfigure[]{\includegraphics[width=0.45\columnwidth,angle=0]{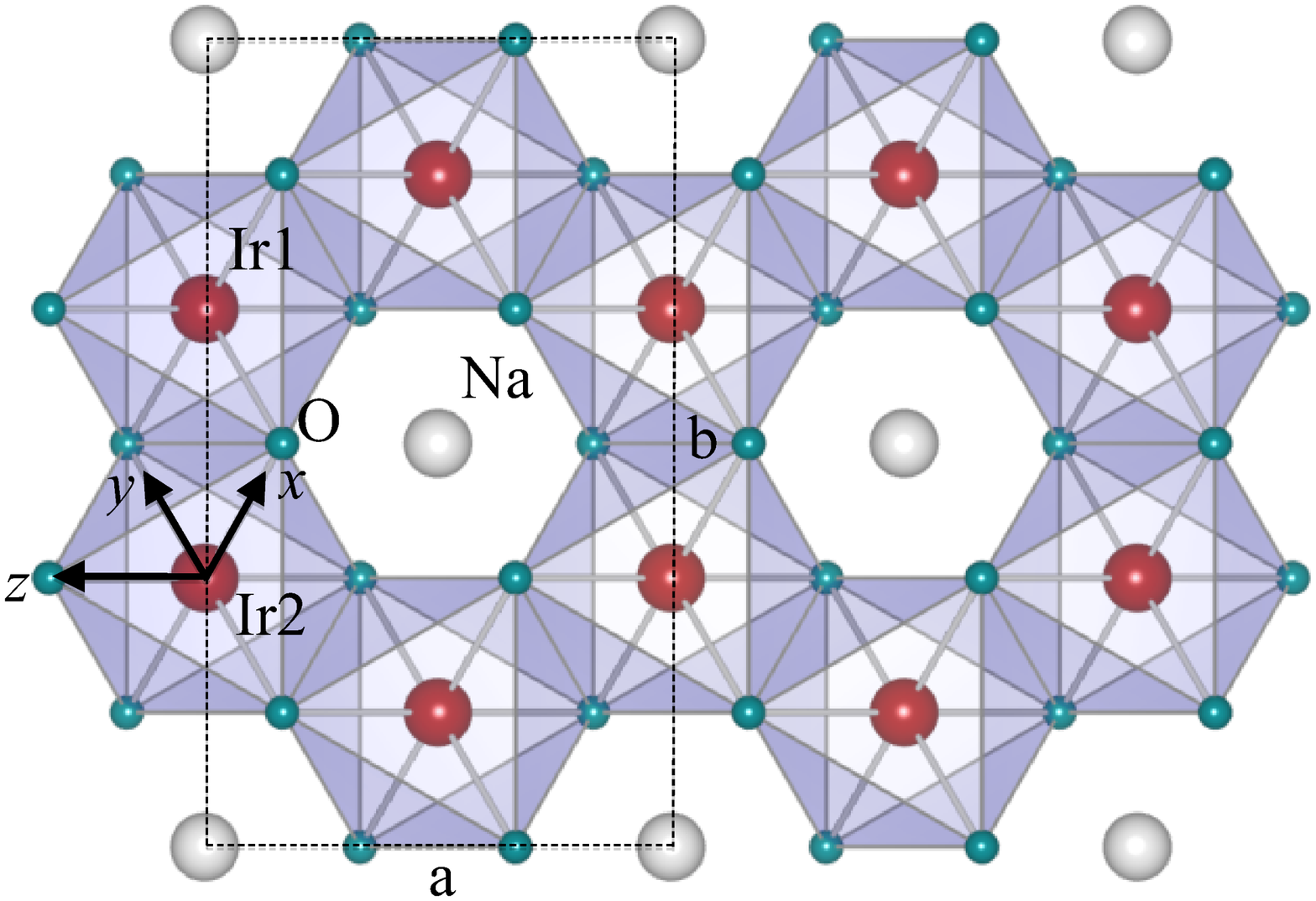}}
\vspace{0.2cm}
\subfigure[]{\includegraphics[width=0.53\columnwidth,angle=0]{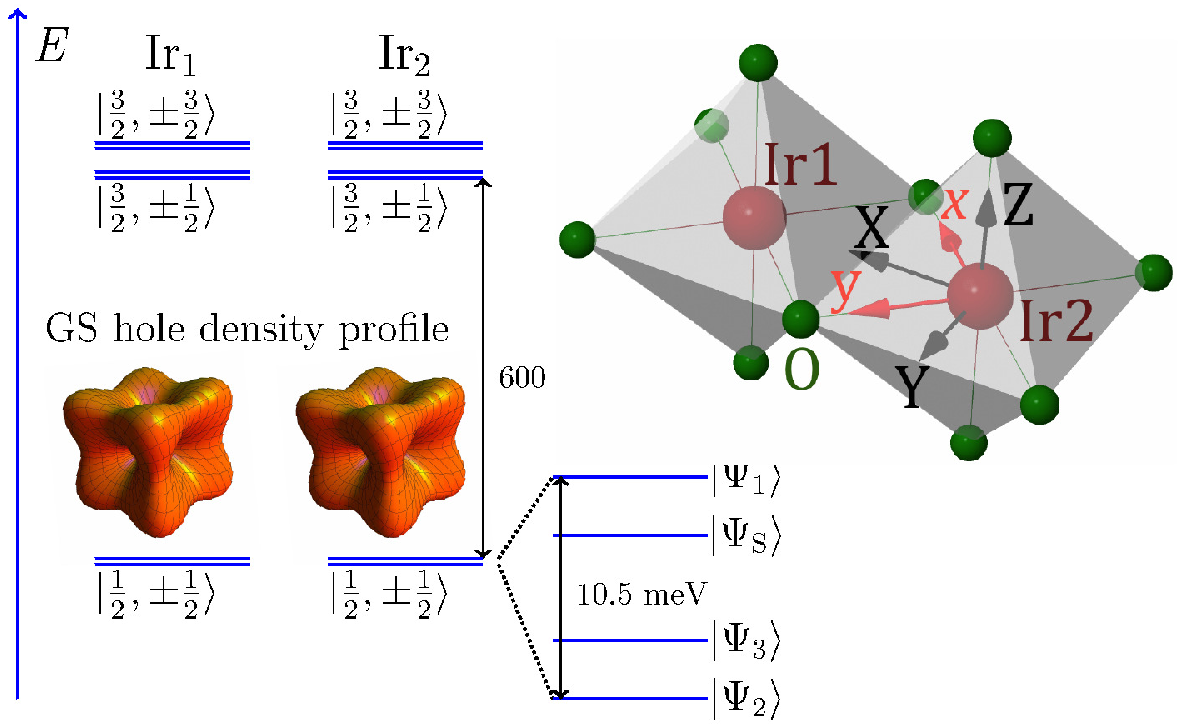}}
\caption{
a) Ir honeycomb layer in Na213, idealized model with cubic IrO$_6$ and NaO$_6$ octahedra of equal size and $90^{\circ}$ Ir-O-Ir bonds.\, b) Low-lying energy levels for two NN octahedra and ground-state (GS) density profiles for the effective spin-orbit $j\!=\!1/2$ states at each Ir site. The $d$-level splittings are not to scale, notations as in Table~\ref{tab:I}. The $(x,y,z)$ coordinate frame used to express the KH Hamiltonian \cite{IrO_kitaev_jackeli_09} is also drawn for one of the Ir-Ir links. For each Ir$_2$O$_2$ plaquette in the actual $C2/m$ structure \cite{Honeycomb_NaIrO_Choi_2012}, due to trigonal squashing of the IrO$_6$ octahedra (normal to the Ir honeycomb plane), the apical-like Ir-O bonds are not along the corresponding Kitaev axis.
}
\label{fig:1}
\end{center}
\end{figure}

\section{Results and discussion}
Multiconfiguration complete-active-space self-consistent-field (CASSCF)
and multireference configuration-interaction (MRCI) calculations~\cite{book_QC_00} were
performed on embedded clusters made of two reference IrO$_6$ octahedra
(for technical details, see Appendix A).
All possible occupations were allowed within the set of $t_{2g}$ orbitals at the two magnetically active
Ir sites in the CASSCF calculations.
The orbitals were optimized for an average of the lowest nine singlets and the nine triplet states.
All these singlet and triplet states entered the spin-orbit calculations, both at the CASSCF
and MRCI levels.
In MRCI, single and double excitations from the Ir $t_{2g}$ shells and the $2p$ orbitals of
the bridging ligands were accounted for.
A similar strategy of explicitly dealing only with selected groups of localized ligand orbitals
was earlier adopted in QC studies on both $3d$ \cite{QC_J_fink94,NOCI_J_oosten96,NOCI_J_hozoi03,J_ligand_calzado03}
and $5d$~\cite{katukuri_PRB_2012,nikolay_cairo_PRB,Os227_bogdanov_12} compounds,
with results in good agreement with the experiment
\cite{katukuri_PRB_2012,nikolay_cairo_PRB,QC_J_fink94,NOCI_J_oosten96,NOCI_J_hozoi03,J_ligand_calzado03}.

\begin{table}[!t]
  \caption{
Energy splittings of the four lowest magnetic states and effective exchange parameters (meV)
for two edge-sharing IrO$_6$ octahedra taken out of the idealized structural model with
$D_{2h}$ bond symmetry.
In a pure KH model, $\Psi_{\mathrm{2}}$ and $\Psi_{\mathrm{3}}$ are degenerate.
$d$(Ir-Ir)=3.133\,\AA  , see text.
}\label{tab:I}
\begin{indented}
\item[]\begin{tabular}{@{}lcc}
\br
Method&CAS+SOC &MRCI+SOC \\
\ns
\mr
\multicolumn{3}{l}{$\measuredangle$(Ir-O-Ir)=90$^{\circ}$\,:}\\
$\Psi_{\mathrm{S}}= (\uparrow\downarrow -\downarrow\uparrow)/\sqrt2$   &$0.0$   &$0.0$      \\
$\Psi_{\mathrm{2}}= (\uparrow\uparrow   +\downarrow\downarrow)/\sqrt2$ &$0.7$   &$0.4$      \\
$\Psi_{\mathrm{3}}= (\uparrow\uparrow   -\downarrow\downarrow)/\sqrt2$ &$0.7$   &$1.1$      \\
$\Psi_{\mathrm{1}}= (\uparrow\downarrow +\downarrow\uparrow)/\sqrt2$   &$1.0$   &$1.1$      \\
($J$,$K$,$D$)                                        &($1.0$,$-0.6$,$0.0$)
                                                                       &($1.1$,$-0.7$,$-0.7$) \\
\ms
\multicolumn{3}{l}{$\measuredangle$(Ir-O-Ir)=98.5$^{\circ}$\,:}\\
$\Psi_{\mathrm{2}}= (\uparrow\uparrow   +\downarrow\downarrow)/\sqrt2$ &$0.0$   &$0.0$      \\
$\Psi_{\mathrm{3}}= (\uparrow\uparrow   -\downarrow\downarrow)/\sqrt2$ &$1.2$   &$2.1$      \\
$\Psi_{\mathrm{S}}= (\uparrow\downarrow -\downarrow\uparrow)/\sqrt2$   &$4.3$   &$5.1$      \\
$\Psi_{\mathrm{1}}= (\uparrow\downarrow +\downarrow\uparrow)/\sqrt2$   &$3.9$   &$6.5$      \\
($J$,$K$,$D$)                                        &($-0.4$,$-6.6$,$-1.2$)
                                                                       &($ 1.4$,$-10.9$,$-2.1$)\\
\br
\end{tabular}
\end{indented}
\end{table}

For a pair $\{i,j\}$ of magnetic sites in systems in which the midpoint of the $ij$ link displays
inversion symmetry, the most general bilinear exchange Hamiltonian is
\be\label{eq:1}
{\mathcal H}_{ij} = J_0\,{\tilde {\bm S}}_i\cdot {\tilde {\bm S}}_j +
                    \displaystyle\sum\limits_{\alpha,\beta\in\{X,Y,Z\}}
                    \Gamma_{\alpha\beta}  {\tilde S}_i^{\alpha} {\tilde S}_j^{\beta}\,,
\ee
where ${\tilde {\bm S}}_i$, ${\tilde {\bm S}}_j$ are pseudospin operators ($\tilde{S}=1/2$) 
\cite{IrO_kitaev_jackeli_09,IrO_kitaev_chaloupka_10}
and the elements $\Gamma_{\alpha\beta}$ form a traceless symmetric second-rank tensor.
It is convenient to choose the $X$ axis along the Ir-Ir link and $Z$ perpendicular to the plaquette
defined by the two Ir ions and the two bridging ligands because in the $C2/m$ crystal structure of Na$_2$IrO$_3$,
for Ir-Ir bonds along $b$, see Fig.~\ref{fig:1}, the Ir-Ir axis is a $C_2$ axis with an orthogonal mirror plane
~\cite{Honeycomb_NaIrO_Choi_2012,Honeycomb_NaIrO_Ye_2012}, i.e., the symmetry of those [Ir$_2$O$_{10}$] units is $C_{2h}$.
With such a choice of the coordinate system only $\Gamma_{YZ} = \Gamma_{ZY}$ are finite and so
\be\label{eq:2}
\bs{\Gamma}=\left(
\begin{array}{ccc}
A & 0 & 0   \\
0 & B & C   \\
0 & C & -A-B
\end{array}
\right)_{\!\!\{X,Y,Z\}}.
\ee
The fact that $Y$ and $Z$ are not $C_2$ axes is related to the
configuration of the four adjacent Ir sites -- two of those are
below and two above the $XY$ plane, with no inversion center -- and the trigonal squashing
of the IrO$_6$ octahedra \cite{Honeycomb_NaIrO_Choi_2012}. The KH
Hamiltonian is however expressed in a $(x,y,z)$ coordinate frame
that has the $(x,y)$ coordinates rotated by $45^{\circ}$ about the
$Z=z$ axis \cite{IrO_kitaev_jackeli_09,IrO_kitaev_chaloupka_10},
as compared to $(X,Y)$, see Fig.~\ref{fig:1}, and ${\bs{\Gamma}}$ then becomes
\be\label{eq:3}
\bs{\Gamma}=\left(
\begin{array}{ccc}
(A+B)/2    &(A-B)/2    &-C/\sqrt{2}   \\
(A-B)/2    &(A+B)/2    &C/\sqrt{2}   \\
-C/\sqrt{2} &C/\sqrt{2} &-A-B
\end{array}
\right)_{\!\!\{x,y,z\}}
\ee
(see Appendix B for details).

For a more transparent picture and better insight into the nature of the NN magnetic couplings,
it is instructive to first consider two-octahedra clusters taken from an idealized crystalline model without trigonal
distortions and with all adjacent Ir and Na sites modeled as identical point charges.
In this case, the overall symmetry is $D_{2h}$ and {\it all} off-diagonal couplings cancel
by symmetry, in the $(X,Y,Z)$ coordinate system with $X$ along the Ir-Ir link.
For an idealized [Ir$_2$O$_{10}$] unit displaying $D_{2h}$ symmetry $C=0$ and the spin
Hamiltonian reduces to
\be\label{eq:4}
{\mathcal H}_{ij}^{\mathrm{D_{2h}}} = J\,{\tilde {\bm S}}_i\cdot {\tilde {\bm S}}_j
                                    + K\,{\tilde S}_i^z {\tilde S}_j^z
                                    + D\left( {\tilde S}_i^x {\tilde S}_j^y + {\tilde S}_i^y {\tilde S}_j^x \right),
\ee
where $K=-\frac{3}{2}(A+B)$, $J = J_0 - K/3$ and $D = \frac{1}{2}(A-B)$.
The off-diagonal $xy$ coupling, last term in (\ref{eq:4}), is allowed by symmetry even for ideal octahedra
at 90$^\circ$ Ir-O-Ir bonding but has been neglected in earlier studies on Na$_2$IrO$_3$
\cite{IrO_kitaev_jackeli_09,IrO_kitaev_chaloupka_10,ZigZag_KH_chalopka_12,Na2IrO3_jkj2j3_kimchi_2011,Mazin_Na213_2013}.

Results of spin-orbit calculations, both at the CASSCF (CAS+SOC) and MRCI (MRCI+SOC)
levels, are listed for idealized [Ir$_2$O$_{10}$] $D_{2h}$ model clusters in Table~\ref{tab:I}.
Such a cluster is highly charged, $12-$.
To ensure charge neutrality, we assigned to each of the 26 adjacent Na and Ir sites
fictitious point charges of $+12/26$.
In the simplest approximation, no farther embedding was used for these $D_{2h}$ clusters.
The {\it ab initio} calculations were performed for both i) regular IrO$_6$ octahedra
and $90^{\circ}$ Ir-O-Ir bond angles and ii) distorted geometries with all ligands
in the $xy$ plane pushed closer to the Ir-Ir axis and therefore larger Ir-O-Ir bond angles but keeping the $D_{2h}$ bond symmetry.
The Ir-Ir distance $d$(Ir-Ir) and in the latter case the Ir-O-Ir angle were set to
3.133 \AA \ and 98.5$^{\circ}$, respectively, average values in the $C2/m$ crystal
structure reported in \cite{Honeycomb_NaIrO_Choi_2012}.

To determine the nature of each spin-orbit state we explicitly compute the dipole and quadrupole
transition matrix elements among those four low-lying states describing the magnetic spectrum
of two edge-sharing octahedra.
A careful symmetry analysis reveals that the spin-orbit wave functions $\Psi_{\mathrm{S}}$, $\Psi_{\mathrm{1}}$,
$\Psi_{\mathrm{2}}$ and $\Psi_{\mathrm{3}}$  defined in Table~\ref{tab:I} transform according to the
$A_g$, $B_{2u}$, $B_{1u}$ and $A_u$ irreducible representations, respectively.
Standard selection rules and the nonzero dipole and quadrupole matrix elements in the QC outputs
then clearly indicate which state is which.
We also carried out the transformation of the spin-orbit wave functions from the
usual \{$L_1$,$M_{L_1}$,$L_2$,$M_{L_2}$,$S$,$M_S$\} basis in standard QC programs to the
\{${\tilde S}_1$,${\tilde S}_2$,${\tilde M}_{S_1}$,${\tilde M}_{S_2}$\}
basis.
This allows the study of $\Psi_{\mathrm{1}}$--$\Psi_{\mathrm{2}}$ mixing when the
point-group symmetry is reduced to $C_{2h}$, see below.
Having the assignment of the states resolved, the $\Psi_{\mathrm{S}}$--$\Psi_{\mathrm{1}}$
splitting provides $J$, the $\Psi_{\mathrm{2}}$--$\Psi_{\mathrm{3}}$
splitting yields $D$, while the difference between the energy of $\Psi_{\mathrm{1}}$
and the average of the $E_2(\Psi_{\mathrm{2}})$ and $E_3(\Psi_{\mathrm{3}})$ eigenvalues equals
$-K/2$, see Appendix B.

The QC data in Table~\ref{tab:I} indicate AF $J$'s, FM $K$'s and off-diagonal
anisotropic couplings comparable in strength to the isotropic $J$ interaction.
Interestingly, the {\it ab initio} MRCI calculations indicate a much stronger $K$ for
nonorthogonal Ir-O-Ir bonds.
This shows that deviations from rectangular geometry on the Ir$_2$O$_2$ plaquette is
{\it not} a negligible factor, as presently assumed in simplified superexchange models for
Kitaev physics in honeycomb Na$_2$IrO$_3$
\cite{IrO_kitaev_jackeli_09,IrO_kitaev_chaloupka_10,ZigZag_KH_chalopka_12}.
The effect of the MRCI treatment is also stronger for nonorthogonal Ir-O-Ir bonds:
the CAS+SOC $K$ and $D$ coupling parameters are enlarged by more
than $50\%$ by including O $2p$ to Ir $5d$ charge-transfer effects, Ir $t_{2g}$ to $e_{g}$
excitations and additional correlations accounted for in MRCI while $J$
changes sign.
This strong enhancement of the FM $K$ for nonorthogonal Ir-O-Ir bonds and the tiny effect
of the MRCI treatment on the FM $K$ for rectangular geometry also disagrees with
predictions of present approximate superexchange models that indicate the $t_{2g}$ to $e_{g}$ excitations
and hopping as the dominant superexchange mechanism, giving rise to a large AF $K$
\cite{ZigZag_KH_chalopka_12}.

\begin{table}
\caption{
Energy splittings of the four lowest magnetic states and effective coupling parameters (meV)
for two NN IrO$_6$ octahedra in the $C2/m$ structure of Ref.~\cite{Honeycomb_NaIrO_Choi_2012}.
The weight of $(\uparrow\downarrow + \downarrow\uparrow)/\sqrt{2}$ and
              $(\uparrow\uparrow   + \downarrow\downarrow)/\sqrt{2}$ in $\Psi_{\mathrm{1}}^{\prime}$
and $\Psi_{\mathrm{2}}^{\prime}$, respectively, is $\approx\!98\%$, see text.
}\label{tab:II}
\begin{indented}
\item[]\begin{tabular}{@{}lcc}
\br
Method&CAS+SOC &MRCI+SOC \\
\ns
\mr
\multicolumn{3}{l}{$\measuredangle$(Ir-O-Ir)=99.45$^{\circ}$, $d$(Ir$_1$-Ir$_2$)=3.138\,\AA \
($\times 1$)$^{\rm a}$\,:}\\
$\Psi_{\mathrm{2}}^{\prime}$                                           &$0.0$    &$0.0$      \\
$\Psi_{\mathrm{3}}= (\uparrow\uparrow  - \downarrow\downarrow)/\sqrt2$ &$0.2$    &$0.5$      \\
$\Psi_{\mathrm{S}}= (\uparrow\downarrow- \downarrow\uparrow)/\sqrt2$   &$4.4$    &$5.5$      \\
$\Psi_{\mathrm{1}}^{\prime}$                                           &$6.3$    &$10.5$     \\
($J$,$K$,$D$)                                        &($1.9$,$-12.4$,$-0.2$) &($5.0$,$-20.5$,$-0.5$)\\
\ms
\multicolumn{3}{l}{$\measuredangle$(Ir-O-Ir)=97.97$^{\circ}$, $d$(Ir$_2$-Ir$_3$)=3.130\,\AA \
($\times 2$)$^{\rm b}$\,:}\\
$\Psi_{\mathrm{2}}^{\prime}$                                           &$0.0$    &$0.0$      \\
$\Psi_{\mathrm{3}}= (\uparrow\uparrow  - \downarrow\downarrow)/\sqrt2$ &$0.3$    &$1.2$      \\
$\Psi_{\mathrm{S}}= (\uparrow\downarrow- \downarrow\uparrow)/\sqrt2$   &$4.6$    &$6.7$      \\
$\Psi_{\mathrm{1}}^{\prime}$                                           &$5.8$    &$8.2$      \\
($J$,$K$,$D$)                                        &($1.2$,$-11.3$,$-0.3$)
                                                                       &($1.5$,$-15.2$,$-1.2$)\\

\br
\end{tabular}
\item[] $^{\rm a}$ $d$(Ir-O$_{1,2}$)=2.056 \AA .
\item[] $^{\rm b}$ $d$(Ir-O$_{1}  $)=2.065 \AA , $d$(Ir-O$_{2}$)=2.083 \AA .
\end{indented}
\end{table}

Relative energies and the resulting effective coupling constants are next given in Table~\ref{tab:II} 
for the experimentally determined $C2/m$ crystal structure of Ref.~\cite{Honeycomb_NaIrO_Choi_2012}.
For this set of calculations we used effective embedding potentials as described in Appendix A.
There are two inequivalent Ir-Ir links in Na$_2$IrO$_3$, displaying different Ir-O-Ir
bond angles and slightly different Ir-O and Ir-Ir distances \cite{Honeycomb_NaIrO_Choi_2012}.
While the [Ir$_2$O$_{10}$] block with larger Ir-O-Ir bond angles (upper part in Table~\ref{tab:II})
displays $C_{2h}$ symmetry, for the other unit of edge-sharing octahedra the point-group
symmetry is even further reduced to $C_i$ (lower part in Table~\ref{tab:II}).
The expressions of the spin-orbit wave functions in the transformed
\{${\tilde S}_1$,${\tilde S}_2$,${\tilde M}_{S_1}$,${\tilde M}_{S_2}$\} basis
show, however, that the mixing of the $\Psi_i$ terms as expressed in the idealized $D_{2h}$
geometry is negligible in the $C2/m$ structure.
Therefore the {\it ab initio} data is mapped also in this case on the effective model described by (\ref{eq:4}).

As for the idealized $D_{2h}$ configuration, the MRCI+SOC results indicate large FM Kitaev
couplings, weaker AF Heisenberg superexchange and sizable $D$ anisotropic interactions.
The latter are not included in the plain KH model 
\cite{IrO_kitaev_jackeli_09,IrO_kitaev_chaloupka_10,ZigZag_KH_chalopka_12,Na2IrO3_jkj2j3_kimchi_2011}
while the signs of $K$ and $J$ that we compute are different from those proposed in the recent
model-Hamiltonian analysis of Ref.~\cite{ZigZag_KH_chalopka_12}.
We note that in agreement with our findings, relatively large FM Kitaev couplings $K$ have been
earlier predicted by Kimchi and You \cite{Na2IrO3_jkj2j3_kimchi_2011} from the analysis of the phase
diagram obtained by ED on modest size clusters and by Foyevtsova {\it et al.} \cite{Mazin_Na213_2013}
on the basis of an effective superexchange model fed with electronic-structure parameters
obtained from density-functional calculations for the same $C2/m$ structure \cite{Honeycomb_NaIrO_Choi_2012}.
However, the NN Heisenberg $J$ is also FM in the latter work, different from the
small AF values we find in the MRCI calculations.
We also find that on each hexagonal Ir$_6$ unit the two Ir-Ir links along the $b$-axis have
effective coupling constants significantly different from the set of parameters associated with
the other four Ir-Ir ``bonds'' due to subtly different oxygen distortions.
Together these findings stress the importance of lattice distortions and symmetry issues
and lay the foundation for rigorous {\it ab initio} investigations of unusually large
anisotropic interactions such as the Kitaev exchange in strongly spin-orbit coupled
$5d$ oxides 
\footnote{
Detailed QC studies of the anisotropic terms have been so far confined to $3d$ oxides,
where the spin-orbit interaction is just a small perturbation and the anisotropic coupling
parameters are orders of magnitude weaker than the isotropic Heisenberg exchange
\cite{DMCuIIMaurice2010,DMCuOPradipto2012,JLiCu2O2Pradipto2012}.
}.

\begin{figure}[b!]
  \centering
\subfigure[]
    {
\includegraphics*[width=0.42\columnwidth,angle=0]{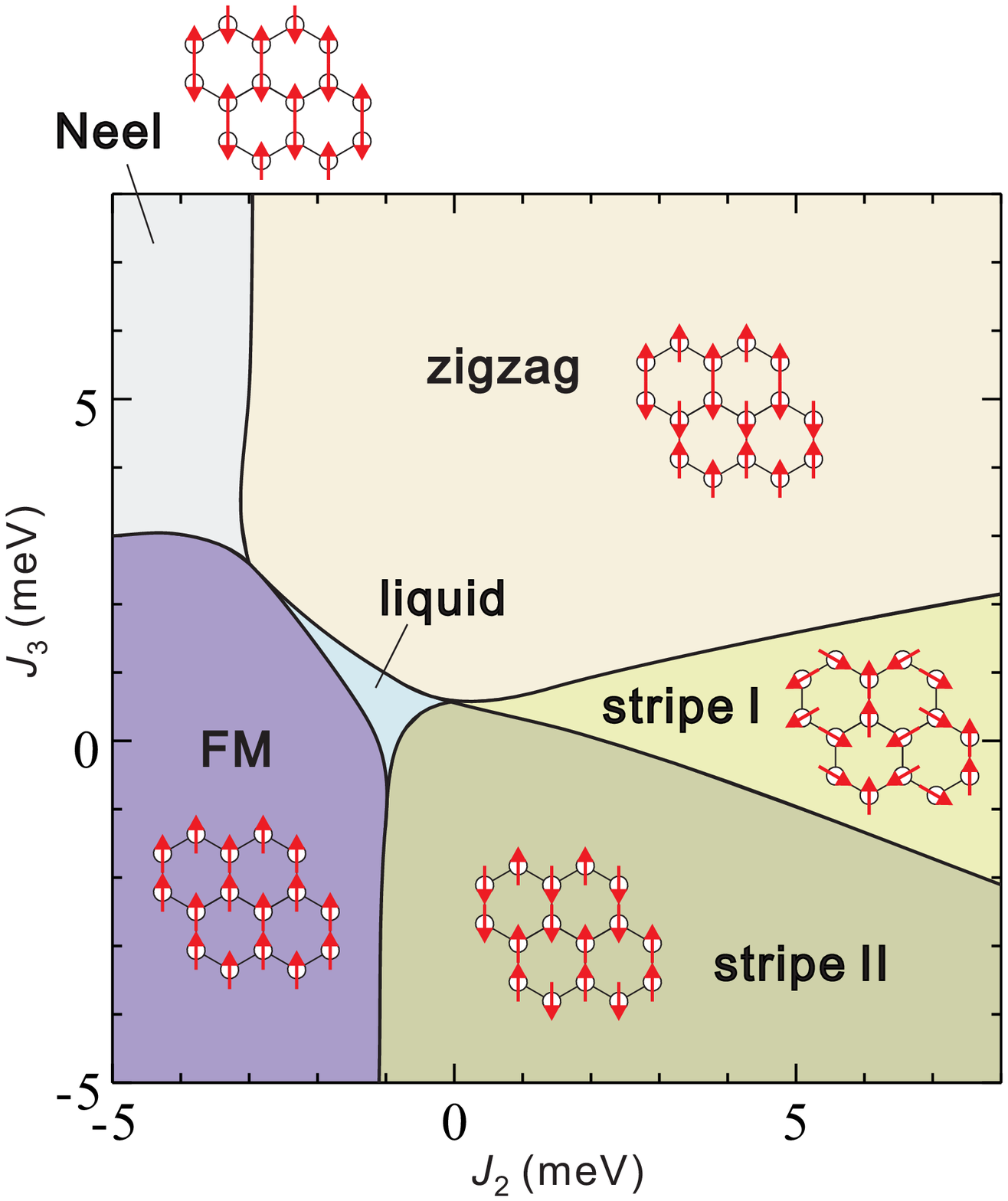}
}
\vspace{-0.5cm}
\subfigure[]{\includegraphics*[width=0.42\columnwidth,angle=0]{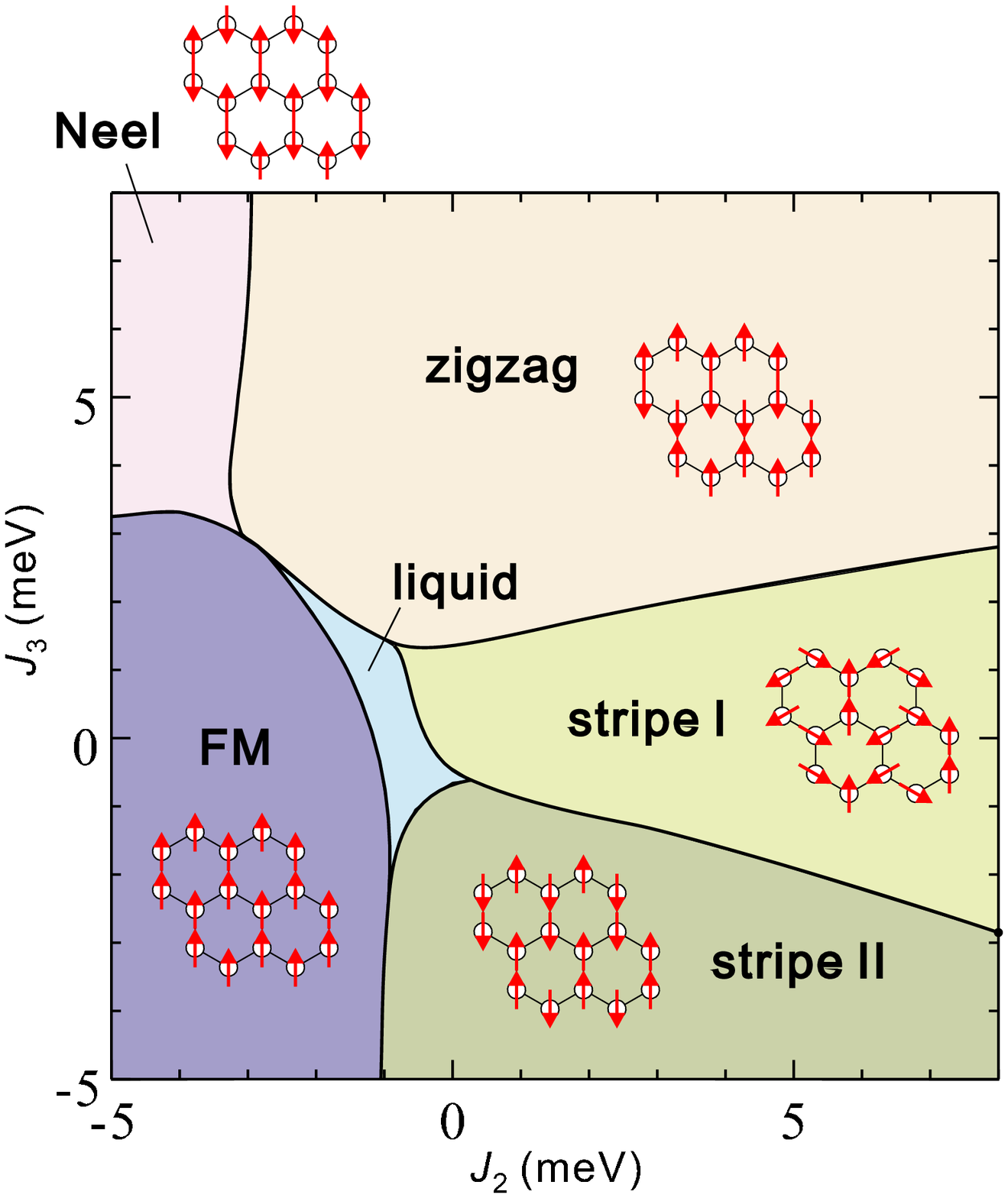}}
\vspace{0.5cm}
\caption{Phase diagram for the effective spin model in (\ref{eq:4})
supplemented by second and third NN couplings $J_2$ and $J_3$,
with $J\!=\!3$~meV, $K\!=\!-17.5$~meV and $D\!=\!0$ (a) or 
$D\!=\!-1$ meV (b), as found by exact diagonalizations on a 24-site cluster.}
\label{fig:2}
\end{figure}

It is known experimentally that Na$_2$IrO$_3$ displays zigzag AF order at low $T$'s
\cite{Honeycomb_NaIrO_Choi_2012,Honeycomb_NaIrO_Ye_2012,Na2IrO3_zigzag_liu_2011}.
It has been also argued that the longer-range magnetic
interactions, up to the second and third Ir coordination shells,
are sizable and AF
\cite{Honeycomb_NaIrO_Choi_2012,Na2IrO3_jkj2j3_kimchi_2011,Na2IrO3_jkj2j3_singh_2012,Mazin_Na213_2013}.
We therefore performed ED calculations for a KH model supplemented
with second and third NN couplings $J_2$ and $J_3$ (see Appendix B), on a 24-site
cluster with periodic boundary conditions as used in earlier
studies \cite{IrO_kitaev_chaloupka_10,ZigZag_KH_chalopka_12}. We
disregarded the presence of two structurally and magnetically
different sets of Ir-Ir links and on the basis of the QC results
of Table~\ref{tab:II}, used on all bonds $J$, $K$ and $D$ coupling
constants of $3$, $-17.5$ and $-1$ meV, respectively (approximately averaged
over all bonds). For a given set of $J_2$ and $J_3$ values
the dominant order is determined according to the wave number
${\bf Q}={\bf Q}_{\rm max}$ giving a maximum value of the static
structure factor $S({\bf Q})$. The resulting phase diagram, see
Fig.~\ref{fig:2}, shows that the zigzag phase is indeed stable in the region
of $J_2$,$J_3$$\gtrsim$2 meV.
We note that positive $J_2$ and $J_3$ values of 4--5 meV would be consistent with the
experimentally observed Curie-Weiss temperature $\approx\!-125$ K
\cite{Na2IrO3_jkj2j3_singh_2012} using $\theta_{\rm
CW}=-\tilde{S}(\tilde{S}+1)(J+2J_2+J_3+K/3)/k_{\rm B}$
\cite{Honeycomb_NaIrO_Choi_2012}.
Thus we propose that an extended spin Hamiltonian based on the nearest-neighbor anisotropic
exchange terms found from the {\em ab initio} QC calculations supplemented by
further-neighbor exchange integrals could provide a realistic starting point to explain
the magnetism of Na$_2$IrO$_3$.

To discuss in more detail the generic phase diagram in Fig.~\ref{fig:2}, we note that
in between the ordered phases we find a narrow spin liquid phase characterized by 
${\bf Q}_{\rm max} \simeq 0$ and very weak spin-spin correlations.
We checked that this phase is adiabatically connected to the Kitaev liquid
phase appearing for larger $|K|$
\cite{IrO_kitaev_chaloupka_10,ZigZag_KH_chalopka_12}. The canted
``stripe I'' phase corresponds to the canted ``phase III'' found
in earlier investigations of the isotropic $J_1$-$J_2$-$J_3$ model
\cite{Rastelli1979,Fouet2001,Alburquerque2011}. 
Interestingly, a finite $D$ enlarges the extent of both the spin liquid and the
stripe I phases, showing that the planar spin fluctuations effectively amplify frustration
in the model.
The ED results show that the phase diagram is also very sensitive to the longer-range
exchange couplings $J_2$ and $J_3$.
These findings are relevant in the context of recent experimental data that indicate a
qualitatively different AF ground state for the related compound Li$_2$IrO$_3$ \cite{Li213_KH_cao_13}.

\section{Conclusions}

In sum, for the honeycomb iridate Na$_2$IrO$_3$, the {\it ab
initio} quantum chemistry calculations show that in a reference
system with $X$ along the Ir-Ir link and $Z$ perpendicular on the
Ir$_2$O$_2$ plaquette the $X$-$Y$ anisotropy is significant and
gives rise in the rotated $(x,y,z)$ Kitaev-type frame
\cite{IrO_kitaev_jackeli_09,IrO_kitaev_chaloupka_10} to
off-diagonal anisotropic terms beyond the plain Kitaev-Heisenberg model.
Nevertheless, the calculations predict that the largest energy scale
is the Kitaev interaction, 10 to 20 meV, while the NN Heisenberg
superexchange and the off-diagonal $xy$ coupling are significantly
weaker.
The quantum chemistry data additionaly establish that the
Kitaev term is FM. Further, all NN couplings are highly sensitive
to subtle distortions involving the O ions. This makes the
material dependence along the Na$_2$IrO$_3$, Li$_2$IrO$_3$ and
Li$_2$RhO$_3$ series an interesting topic for future
investigations.
Large variations of the effective couplings as
function of bond lengths and bond angles and a variable degree of
``inequivalence'' of those sets of parameters for structurally
distinct Ir-Ir links in the honeycomb layer may in principle give
rise to very different types of magnetic ground states in
different $5d^5$ or $4d^5$ honeycomb compounds. Strong exchange
anisotropy is also a topic of active research in the field of
molecular magnetism and single-molecule magnets. The focus there
has so far been on $3d$ and $4d$ compounds
\cite{aniso_chibotaru_2003,aniso_maurice_2010,aniso_tsukerblat_2011}
but clearly $5d$ systems with stronger spin-orbit couplings may
provide new playgrounds in this research area too.

After submission of our paper, results of ED calculations including off-diagonal $D$ type terms were also reported in Ref.~\cite{kee}.

\section{Acknowledgements}

We thank  N.~A.~Bogdanov, G.~Khaliullin, D.~I.~Khomskii, H.~Gretarsson, Y.-J.~Kim, H.~Stoll and P.~Fulde for insightful discussions. L.~H. acknowledges financial support from the German Research Foundation (Deutsche Forschungsgemeinschaft, DFG).

\section{Appendix A: Computational details in the QC calculations}

For the computation of the intersite spin couplings, two reference NN IrO$_6$ octahedral units are considered.
Since it is important to describe the finite charge distribution at sites in the immediate neighborhood
\cite{CuO2_dd_hozoi11,qc_NNs_degraaf_99},
the closest 22 Na neighbors and the other four, adjacent octahedra are also explicitly
included in the actual cluster.
However, to make the analysis of the low-lying magnetic states tractable, we cut off
the spin couplings with the adjacent $5d$ ions by replacing those open-shell Ir$^{4+}$ $5d^5$
NN's with closed-shell Pt$^{4+}$ $5d^6$ species.
This is an usual procedure in quantum chemistry studies on transition-metal systems, see, e.g.,
Refs.~\cite{Na2V2O5_hozoi_02,katukuri_PRB_2012,nikolay_cairo_PRB,Ir227_hozoi_12,Os227_bogdanov_12,qc_NNs_degraaf_99,SIA_Fe_maurice_2013},
and here allows a straightforward mapping of the {\it ab initio} data onto the effective spin model.
It has been shown in earlier work that this way of modeling the NN $5d$ ions does not affect the size
of the $t_{2g}$ splittings at the central, reference site \cite{katukuri_PRB_2012,nikolay_cairo_PRB}
and that the computed Heisenberg couplings agree well with estimates derived from experiment
\cite{katukuri_PRB_2012,nikolay_cairo_PRB}.
The surrounding solid-state matrix is described as a finite array of point charges fitted to reproduce the
crystal Madelung field in the cluster region.
Most of the {\it ab initio} calculations were carried out with the {\sc molpro} quantum chemistry package
\cite{molpro_brief}.
Test calculations in finite magnetic fields were also performed with the {\sc orca} program \cite{orca}.

We used energy-consistent relativistic pseudopotentials with quadruple-zeta basis sets for the valence shells
of the two reference Ir ions \cite{ECP_Stoll_2},
all-electron quintuple-zeta basis sets for the bridging ligands \cite{GBas_molpro_2p} and
triple-zeta basis functions for the other O's of the two reference octahedra \cite{GBas_molpro_2p}.
For the NN Pt$^{4+}$ $5d^6$ species triple-zeta basis sets were applied \cite{ECP_Stoll_2} while the
other O ions not shared with the central octahedra were modeled with minimal atomic-natural-orbital basis functions
\cite{ANOs_pierloot_95}.
All occupied shells at the NN Na$^{+}$ sites were incorporated in the large-core pseudopotentials
and each of the Na $3s$ orbitals was described with a single basis function \cite{ECP_Na_Stollgroup}.
For the central Ir ions and the two bridging ligands we also employed polarization functions,
two Ir $f$ and four $d$ O functions \cite{ECP_Stoll_2,GBas_molpro_2p}.
To separate the metal $5d$ and O $2p$ valence orbitals into different groups, we used the orbital localization module
available in {\sc molpro}.
The MRCI calculations were carried out for each spin multiplicity, singlet or triplet, as nine-root
calculations.
Only the four low-lying spin-orbit states are relevant for the analysis of the NN magnetic
interactions.
The higher-lying spin-orbit states imply an excitation energy of at least 0.6 eV.
This gap concerns the $j\!=\!1/2$ to $j\!=\!3/2$ transitions \cite{213_rixs_gretarsson_2012}.

The calculations in finite magnetic fields performed with the {\sc orca} package were used to crosscheck the assignment of the lowest
four magnetic states made on the basis of the dipole/quadrupole transition matrix elements,
symmetry analysis and selection rules.
These test calculations were carried out only for idealized $D_{2h}$ structural models such as those of Table~\ref{tab:I}.
It can be shown that when $\mathbf{B}\!\parallel\!\mathrm{O}z$, $\mathbf{B}\!\parallel\!\mathrm{O}y$
or $\mathbf{B}\!\parallel\!\mathrm{O}x$, state $\Psi_{\mathrm{1}}$, $\Psi_{\mathrm{2}}$ or
$\Psi_{\mathrm{3}}$, respectively, defined as in Table~\ref{tab:I}, should not change energy.
This provides a quick alternative way of identifying the low-lying spin-orbit states.

\section{Appendix B: Lattice spin model }

\begin{figure}[!t]
  \begin{center}
\includegraphics[width=0.50\linewidth]{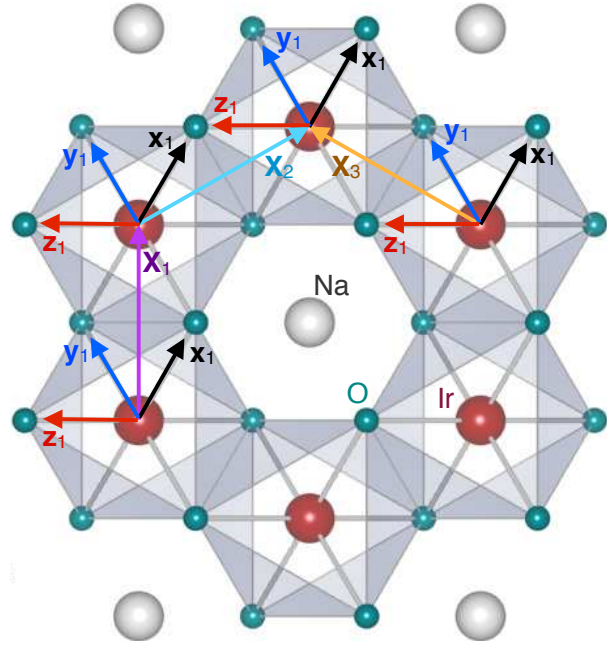}
\caption{
The two types of local reference frames, $\{\vec{X}_b,\vec{Y}_b,\vec{Z}_b\}$ and $\{\vec{x}_b,\vec{y}_b,\vec{z}_b\}$, that are introduced for each of the three different types of bonds on the honeycomb layer.
}
\label{fig:localframes}
\end{center}
\end{figure}

In the main body of this article, we made use of two different local reference frames,
namely \{$\vec{x}_b$,$\vec{y}_b$,$\vec{z}_b$\} and
\{$\vec{X}_b$,$\vec{Y}_b$,$\vec{Z}_b$\} =$\{\frac{\vec{x}_b+\vec{y}_b}{\sqrt{2}},\frac{-\vec{x}_b+\vec{y}_b}{\sqrt{2}},\vec{z}_b\}$,
for each of the three bond types $b\!=\!1\!-\!3$, see Fig.~\ref{fig:localframes}.
Choosing the global frame to be $\{\vec{x},\vec{y},\vec{z}\}\equiv\{\vec{x}_1,\vec{y}_1,\vec{z}_1\}$,
the local frames $\{\vec{X}_b,\vec{Y}_b,\vec{Z}_b\}$ are expressed as
\bea\label{eq:frames}
&&\{\vec{X}_1,\vec{Y}_1,\vec{Z}_1\}=\{\frac{\vec{x}+\vec{y}}{\sqrt{2}}, \frac{-\vec{x}+\vec{y}}{\sqrt{2}}, \vec{z}\}, \nonumber\\
&&\{\vec{X}_2,\vec{Y}_2,\vec{Z}_2\}=\{\frac{\vec{x}-\vec{z}}{\sqrt{2}}, -\frac{\vec{x}+\vec{z}}{\sqrt{2}}, \vec{y}\}, \\
&&\{\vec{X}_3,\vec{Y}_3,\vec{Z}_3\}=\{ \frac{\vec{y}+\vec{z}}{\sqrt{2}}, \frac{-\vec{y}+\vec{z}}{\sqrt{2}}, \vec{x}\}~. \nonumber
\eea
We note that the system is invariant under a two-fold rotation around $\vec{X}_1$ and
that the directions of $\vec{X}_2$ and $\vec{X}_3$ are chosen to map to each other under
this rotation (otherwise the sign of the exchange coupling parameter $C$ is arbitrary).

To be explicit, let us write down all NN interaction terms for each of the three types of
bonds $b$ in the corresponding reference frame $\{\vec{X}_b,\vec{Y}_b,\vec{Z}_b\}$:
\bea
\fl
\mc{H}_{\langle ij\rangle \in b} &=& J_{0,b} \vec{\tilde S}_i\cdot\vec{\tilde S}_j + A_b {\tilde S}_i^{X_b}{\tilde S}_j^{X_b} + B_b {\tilde S}_i^{Y_b}{\tilde S}_j^{Y_b} -(A_b+B_b) {\tilde S}_i^{Z_b}{\tilde S}_j^{Z_b} + C_b ({\tilde S}_i^{Y_b}{\tilde S}_j^{Z_b}+{\tilde S}_i^{Z_b}{\tilde S}_j^{Y_b})\nonumber\\
\fl
&=&J_b\vec{\tilde S}_i\cdot\vec{\tilde S}_j + K_b {\tilde S}_i^{Z_b}{\tilde S}_j^{Z_b} + D_b ({\tilde S}_i^{X_b}{\tilde S}_j^{X_b}-{\tilde S}_i^{Y_b}{\tilde S}_j^{Y_b}) + C_b ({\tilde S}_i^{Y_b}{\tilde S}_j^{Z_b}+{\tilde S}_i^{Z_b}{\tilde S}_j^{Y_b})\,, \label{eq:XYZ}
\eea
where $J_b=J_{0,b}+\frac{1}{2}(A_b+B_b)$, $K_b=-\frac{3}{2}(A_b+B_b)$ and
$D_b=\frac{1}{2}(A_b-B_b)$.
We can rewrite these terms in the global frame $\{\vec{x},\vec{y},\vec{z}\}$ using (\ref{eq:frames}):
\bea
\fl
\mc{H}_{\langle ij\rangle \in b=1} = J_1 \vec{\tilde S}_i\cdot\vec{\tilde S}_j + K_1 {\tilde S}_i^{z}{\tilde S}_j^{z} + D_1 ( {\tilde S}_i^{x}{\tilde S}_j^{y}+{\tilde S}_i^{y}{\tilde S}_j^{x}) \nonumber\\
+ \frac{C_1}{\sqrt{2}} ({\tilde S}_i^{y}{\tilde S}_j^{z}+{\tilde S}_i^{z}{\tilde S}_j^{y}-{\tilde S}_i^{x}{\tilde S}_j^{z}-{\tilde S}_i^{z}{\tilde S}_j^{x})\,, \nonumber\\
\fl
\mc{H}_{\langle ij\rangle \in b=2} = J_2  \vec{\tilde S}_i\cdot\vec{\tilde S}_j + K_2  {\tilde S}_i^{y}{\tilde S}_j^{y} + D_2(-{\tilde S}_i^{x}{\tilde S}_j^{z}-{\tilde S}_i^{z}{\tilde S}_j^{x}) \nonumber\\
+\frac{C_2}{\sqrt{2}} (-{\tilde S}_i^{x}{\tilde S}_j^{y}-{\tilde S}_i^{y}{\tilde S}_j^{x}-{\tilde S}_i^{z}{\tilde S}_j^{y}-{\tilde S}_i^{y}{\tilde S}_j^{z})\,, \nonumber\\
\fl
\mc{H}_{\langle ij\rangle \in b=3} = J_3  \vec{\tilde S}_i\cdot\vec{\tilde S}_j + K_3  {\tilde S}_i^{x}{\tilde S}_j^{x} + D_3 ({\tilde S}_i^{y}{\tilde S}_j^{z}+{\tilde S}_i^{z}{\tilde S}_j^{y}) \nonumber\\
+{ \frac{C_3}{\sqrt{2}} (-{\tilde S}_i^{x}{\tilde S}_j^{y}-{\tilde S}_i^{y}{\tilde S}_j^{x}+{\tilde S}_i^{x}{\tilde S}_j^{z}+{\tilde S}_i^{z}{\tilde S}_j^{x})}\,. 
\eea
For Na$_2$IrO$_3$ bonds 2 and 3 are equivalent and therefore $J_2=J_3$, $K_2=K_3$, $D_2=D_3$, $C_2=C_3$.

In $C_{2h}$ symmetry, $D_b\!=\!0$ and for a given bond $b$ the eigenvalues of the Hamiltonian defined by Eqn. \ref{eq:XYZ} are 
\begin{subequations}
\begin{align}
E_{\mathrm{S}}=-\frac{3J_{0,b}}{4}\,,\\
E_1^{\prime}  =\frac{J_{0,b}+A_b+\sqrt{(A_b+2B_b)^2+4C_b^2}}{4}\,,\\
E_2^{\prime}  =\frac{J_{0,b}+A_b-\sqrt{(A_b+2B_b)^2+4C_b^2}}{4}\,,\\
E_3           =\frac{J_{0,b}-2A_b}{4}\,.
\end{align}
\end{subequations}
These expressions become more complicated for point-group symmetries lower than $C_{2h}$.
The analysis of the spin-orbit wavefunctions in the transformed
\{${\tilde S}_1$,${\tilde S}_2$,${\tilde M}_{S_1}$,${\tilde M}_{S_2}$\} basis shows, however, that
the mixing of the $D_{2h}$ $\Psi_i$ eigenvectors for distorted clusters and lower
symmetries of the [Ir$_2$O$_{10}$] blocks (see Table~\ref{tab:II}) is negligible in the $C2/m$ structure
determined by Choi {\it et al.} \cite{Honeycomb_NaIrO_Choi_2012}.

For $D_{2h}$ symmetry (see Table~\ref{tab:I}), 
$C_b=0$ and the eigenvalues in the effective 2-site, 4-state problem are
$E_{\mathrm{S}} = -3J_{0,b}/4$,
$E_1 = (J_{0,b}+2A_b+2B_b)/4$,
$E_2 = (J_{0,b}-2B_b)/4$ and
$E_3 = (J_{0,b}-2A_b)/4$.

\section*{References}
\bibliographystyle{iopart-num}
\bibliography{Na213}

\end{document}